\documentclass[twocolumn]{article}
\usepackage[utf8]{inputenc}
\usepackage{graphicx}

\usepackage{amssymb}
\usepackage{amsmath}

\usepackage{caption}
\usepackage{subcaption}
\usepackage{algorithm}
\usepackage{algorithmic}

\begin{document}

\title{Technical note: Hybrid Loewner Data Driven Control}
\author{Pierre Vuillemin$^\dagger$, Pauline Kergus$^\dagger$ and Charles Poussot-Vassal$^\dagger$\\
{\small
$^\dagger$ONERA / DTIS, Universit\'e de Toulouse, F-31055 Toulouse, France
}\\
{\small \texttt{pierre.vuillemin@onera.fr}, \texttt{pauline.kergus@onera.fr}, \texttt{charles.poussot-vassal@onera.fr}
}}

\maketitle

\begin{abstract}

This note describes how the Loewner framework can be exploited to create a discrete-time control-law from frequency-data of a continuous-time plant so that their hybrid interconnection matches a given continuous-time reference model up to the Nyquist frequency. The resulting Hybrid Loewner Data Driven Control scheme is illustrated on two numerical examples.

\end{abstract}

\section{Introduction}

Data-Driven Control (DDC) methods (see e.g. \cite{hou2013model} for an overview) enable to create a control-law solely based on input-output data without requiring to explicitly identify a model for the plant first. More specifically, considering some reference model that the closed-loop should match, it is then possible to express the input-output data that the ideal controller should produce. The latter can then be identified. DDC methods distinguish themselves depending on the nature of the data and the structure of the control-law. This note only deals with the Linear Time Invariant (LTI) case and frequency-domain data.

The DDC framework is generally considered either completely in continuous-time or completely in discrete-time. However, it is quite common in real-world control applications that the phenomenon to be controlled is known through continuous-time data while the control-law will eventually be implemented digitally on a computer. This is generally dealt with after the synthesis by discretising the control-law. In this note, we show how some DDC methods can readily be adapted to account for the hybrid nature of the interconnection thus directly embedding the discretisation step.

For that purpose, the idea presented in \cite{vuillemin2019discretisation} for the discretisation of a LTI dynamical model is coupled with the Loewner-based DDC (LDDC) method developed in \cite{kergus:2019:phd}. The next section details the approach and highlights the main differences with the original LDDC method.

\section{Hybrid Loewner Data Driven Control}

\begin{figure}
    \centering
      \begin{subfigure}[b]{0.8\linewidth}
    \includegraphics[width=\linewidth]{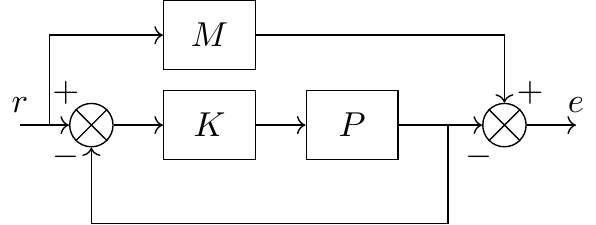}
    \caption{Continuous-time case}
    \label{fig:ddc}
    \end{subfigure}
    
      \begin{subfigure}[b]{\linewidth}
    \includegraphics[width=\linewidth]{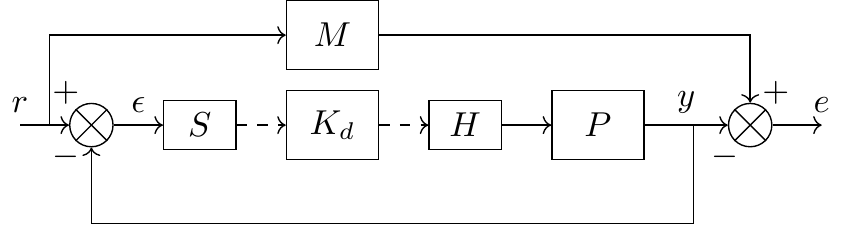}
    \caption{Hybrid case}
    \label{fig:hddc}
    \end{subfigure}
    \caption{Scheme for model-reference control.}
    \label{fig:ddcscheme}
\end{figure}

Let us consider a Single Input Single Output (SISO) LTI plant $P$ and a given reference model $M$. With reference to Figure \ref{fig:ddc}, model-reference control consists in finding a control-law $K$ that minimises the mismatch between the closed-loop and the reference model, i.e. the transfer from $r$ to $e$. If the plant $P$ is known, then the ideal control $K^\star$ is given as 
\begin{equation}
    K^\star = P^{-1}M(I - M)^{-1}.
    \label{eq:kstar}
\end{equation}
However when $P$ is solely known through input-output data, $K^\star$ cannot be obtained as in \eqref{eq:kstar}. Instead, the idea in DDC consists in identifying it from its input-output data.

Assume $P$ is known through frequency-domain data $\{ \omega_i, \Phi_i\}_{i=1}^n$ with $\Phi_i = P(j\omega_i)$, the L-DDC method \cite{kergus:2019:phd} exploits the Loewner approach \cite{Mayo:2007} to build a control-law $K$ that matches the frequency response of the ideal one at $\omega_i$, i.e.
\begin{equation}
    K(j\omega_i) = K^\star(j\omega_i) = \Phi_i^{-1} M(j\omega_i) (I - M(j\omega_i))^{-1}.
    \label{eq:kstarjw}
\end{equation}

If instead of a continuous-time controller $K$, a discrete-time one $K_d$ with a sampling period $T$ is sought, then the scheme \ref{fig:ddc} must be completed by analog/digital converters as in Figure \ref{fig:hddc} where $S$ and $H$ are the ideal sampler and holder (see e.g. \cite[chap.3]{chen1995optimal}). Such a mixed discrete/continuous interconnection is called a \emph{sampled-data system} (see \cite{chen1995optimal} and references therein for an overview) which requires dedicated tools to be studied. In particular, even if $P$, $M$ and $K_d$ are LTI, the overall interconnection is not. It is instead a $T$-periodic system\footnote{A system $G$ is $T$-periodic if $D_T G = G D_T$ where $D_T$ is time-delay of length $T$. A LTI model is $T$-periodic for all $T$.} and has no transfer function thus preventing from using the equation \eqref{eq:kstarjw} as in standard DDC. 

However it is possible to express the relation between the Fourier transforms of $r$ and $e$. In particular, using the frequency-domain relations for $S$ and $H$ detailed in  \cite[chap.3]{chen1995optimal}, the relation between the input $\epsilon$ of the sampler and the output $y$ of the plant is 
\begin{equation}
\hat{y}(j\omega) = \underbrace{P(j\omega) R(j\omega) K_d(e^{j\omega T})}_{F(j\omega)} \sum_{k\in\mathbb{Z}} \hat{\epsilon}(j\omega + jk\omega_s)
\end{equation}
where $R(s) = \frac{1 - e^{-sh}}{sh}$ and $\omega_s=2 \pi /T$ is the sampling frequency. Assuming that $\hat{\epsilon}$ is bandlimited\footnote{For instance, the sampler can be completed with an anti-aliasing filter.} on $[-\omega_N, \omega_N]$, $\omega_N = \omega_s/2$, then for $|\omega|< \omega_N$, the usual feedback relation is retrieved,
\begin{equation}
    \hat{y}(j\omega) = (I + F(j\omega))^{-1} F(j\omega) \hat{r}(j\omega).
\end{equation}
Therefore, the mismatch error with the reference model satisfies, for $|\omega|< \omega_N$, 
\begin{equation}
    \hat{e}(j\omega) = \left (M(j\omega) - (I + F(j\omega))^{-1} F(j\omega) \right) \hat{r}(j\omega).
\end{equation}
To minimise this mismatch, the ideal discrete-time control-law $K_d^\star$ should be such that for $|\omega|< \omega_N$,
\begin{equation}
    M(j\omega) = (I + F(j\omega))^{-1} F(j\omega),
\end{equation}
or equivalently,
\begin{equation}
    K_d^\star(e^{j\omega T}) =  (P(j\omega) R(j\omega))^{-1} M(j\omega)(I - M(j\omega))^{-1}.
    \label{eq:kdstar}
\end{equation}
By sampling the interval $[0,\omega_N]$, this infinite number of interpolation conditions is approximated by a finite number of interpolations conditions that our control-law $K_d$ should satisfy,
\begin{equation}
\small
    K_d(e^{j\omega_iT})  = (\Phi_i R(j\omega_i))^{-1} M(j\omega_i)(I - M(j\omega_i))^{-1}.
    \label{eq:kdjwinterp}
\end{equation}
Such an interpolant can easily be created with the Loewner approach as illustrated in  \cite{vuillemin2019discretisation} for the discretisation objective. 

The two main differences with the standard DDC framework lie in the fact that the frequency response of the plant is filtered by the transfer function of the holder and that the control-law must match the data on the unit circle instead of the imaginary axis. The overall approach is summarised in Algorithm \ref{theAlg} in its simple form. The following remarks can be made:
\begin{itemize}
    \item The Multiple Input Multiple Output (MIMO) case can be handled as in the L-DDC approach by completing the interpolation conditions \eqref{eq:kdjwinterp} with tangential directions to fit the Loewner framework.
    \item If the stability of the resulting controller $K_d$ matters, then an additional projection onto a stable subspace (see e.g. \cite{mari:modifications:2000}) may be performed. During this step, the error induced to the interpolation conditions \eqref{eq:kdjwinterp} should be monitored to give insights on how well the reference model is matched.
    \item In Algorithm \ref{theAlg}, the order of the resulting control-law $K_d$ is determined by the Loewner approach which may thus results in a large dimension. In that case, an additional reduction step can be used. As in the previous point, the interpolation error must also be monitored during this step.
    \item Stability of the overall hybrid interconnection \ref{fig:hddc} is not guaranteed and should be checked a posteriori with dedicated sampled-data systems theory tools \cite{chen1995optimal}.
\end{itemize}

\begin{algorithm}[t]
\caption{Hybrid L-DDC (full-order SISO case)} \label{theAlg}
\begin{algorithmic}[1]
\REQUIRE A sampling period $T>0$, data of the plant $\{\omega_i, \Phi_i\}_{i=1}^{N}$ sampled within $[0, \omega_N]$, a reference model $M$
\STATE Compute the frequency response $\Psi_i = K_d^\star(e^{j\omega_i T})$ of the ideal discrete-time control-law based on equation \eqref{eq:kdjwinterp}.
\STATE Apply the Loewner approach to the data set $\{ e^{j\omega_i T}, \Psi_i\}_{i=1}^N$ to obtain $K_d$
\RETURN the discrete-time control law $K_d$.
\end{algorithmic}
\end{algorithm}

\section{Numerical illustration}

This section shows what may be achieved with HLDDC on a DC motor model and a flexible transmission.

\subsection{DC motor}
Let us consider the following plant model,
\begin{equation}
    P(s) = \frac{0.01}{0.005s^2 + 0.06s +0.1001}
\end{equation}
for which we would like to design a control-law so that the closed-loop behaves as a fully damped second-order model with unitary static gain,
\begin{equation}
    M(s) = \frac{1}{s^2 + 2s + 1}
\end{equation}

In this simple case, the closed-loop is achievable and the L-DDC approach \cite{kergus:2019:phd} enables to retrieve\footnote{with $100$ samples logarithmically spaced between $0.1$ and $10^3$.} the ideal control-law $K^\star$ exactly,
\begin{equation}
    K^\star(s) = \frac{0.5s^2 + 6s + 10.01}{s^2 + 2s}
\end{equation}

The control-law $K^\star$ is discretised with the Tustin approach for the sampling period $T = 0.9s$ leading to $K_d^{tus}$,
\begin{equation}
K_d^{tus}(z) = \frac{2.751 z^2 + 1.607 z - 0.09104}{z^2 - 1.053 z + 0.05263}.
\end{equation}
The latter is compared to the control-law $K_d$ obtained with Algorithm \eqref{theAlg} considered for $50$ frequency samples logarithmically spaced between $0.1$ and $0.95 \omega_N$, completed with a projection onto a stable subspace and a reduction step to the same order as $K_d^{tus}$. The resulting controller is,
\begin{equation}
K_d(z) =  \frac{4.244 z^2 + 3.945 z + 0.2366}{z^2 - 0.1295 z - 0.8705}
\end{equation}

\begin{figure}
    \centering
    \includegraphics[width=\linewidth]{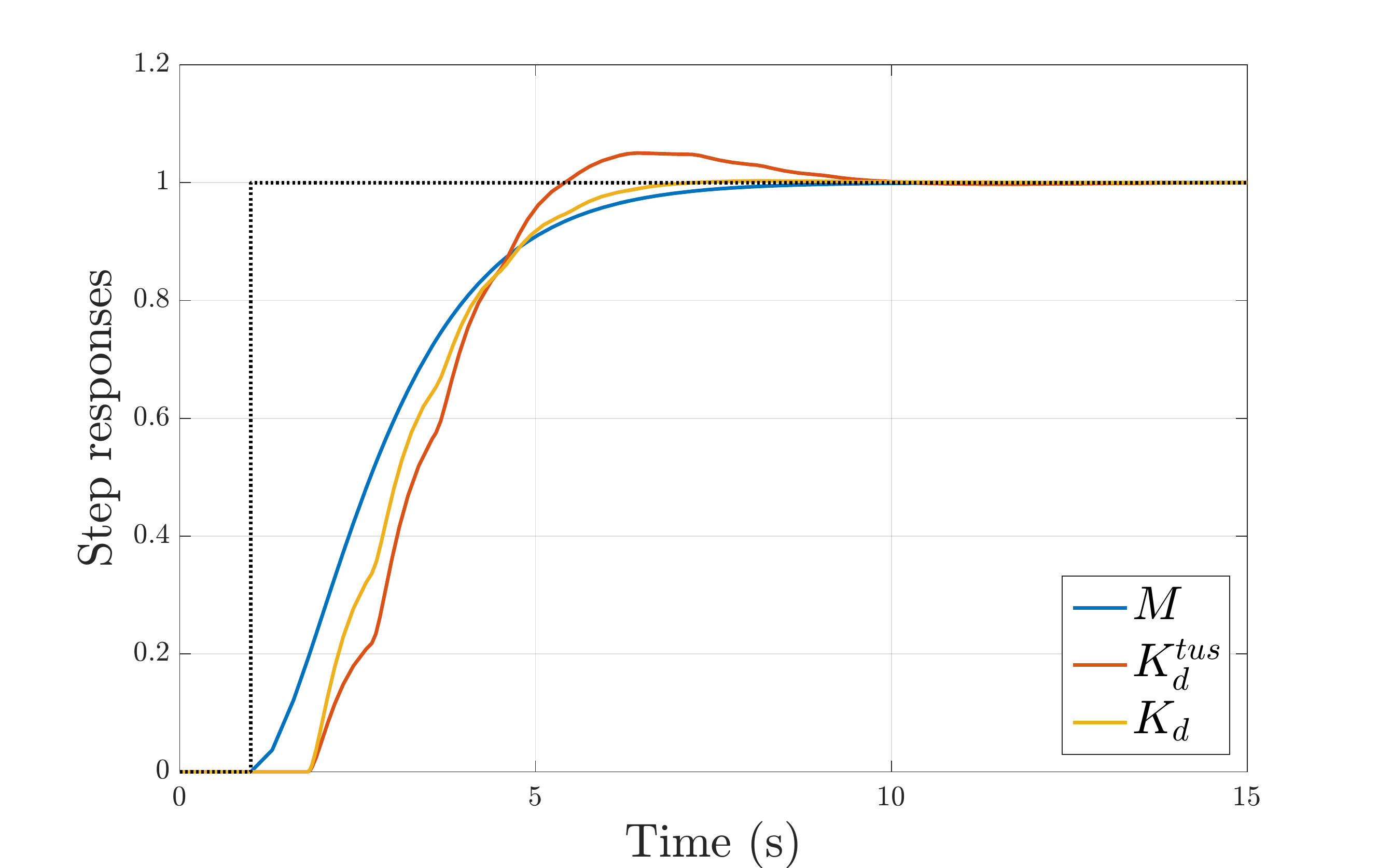}
    \caption{Step responses of the reference model and the closed-loops obtained with the direct approach and the separated discretisation of the control-law for the DC-motor.}
    \label{fig:dcmot}
\end{figure}

The step responses of the reference model and the closed-loops obtained with $K_d^{tus}$ and $K_d$ are plotted in Figure \ref{fig:dcmot}. Due to the large sampling period $T$, both discrete-time control-law achieve a degraded behaviour in comparison to $M$. While $K_d^{tus}$ leads to overshoot in the response and a higher settling time, the controller $K_d$ manages to maintain a closer response to the reference model without overshoot and with a comparable settling time.

\begin{figure}
    \centering
    \includegraphics[width=\linewidth]{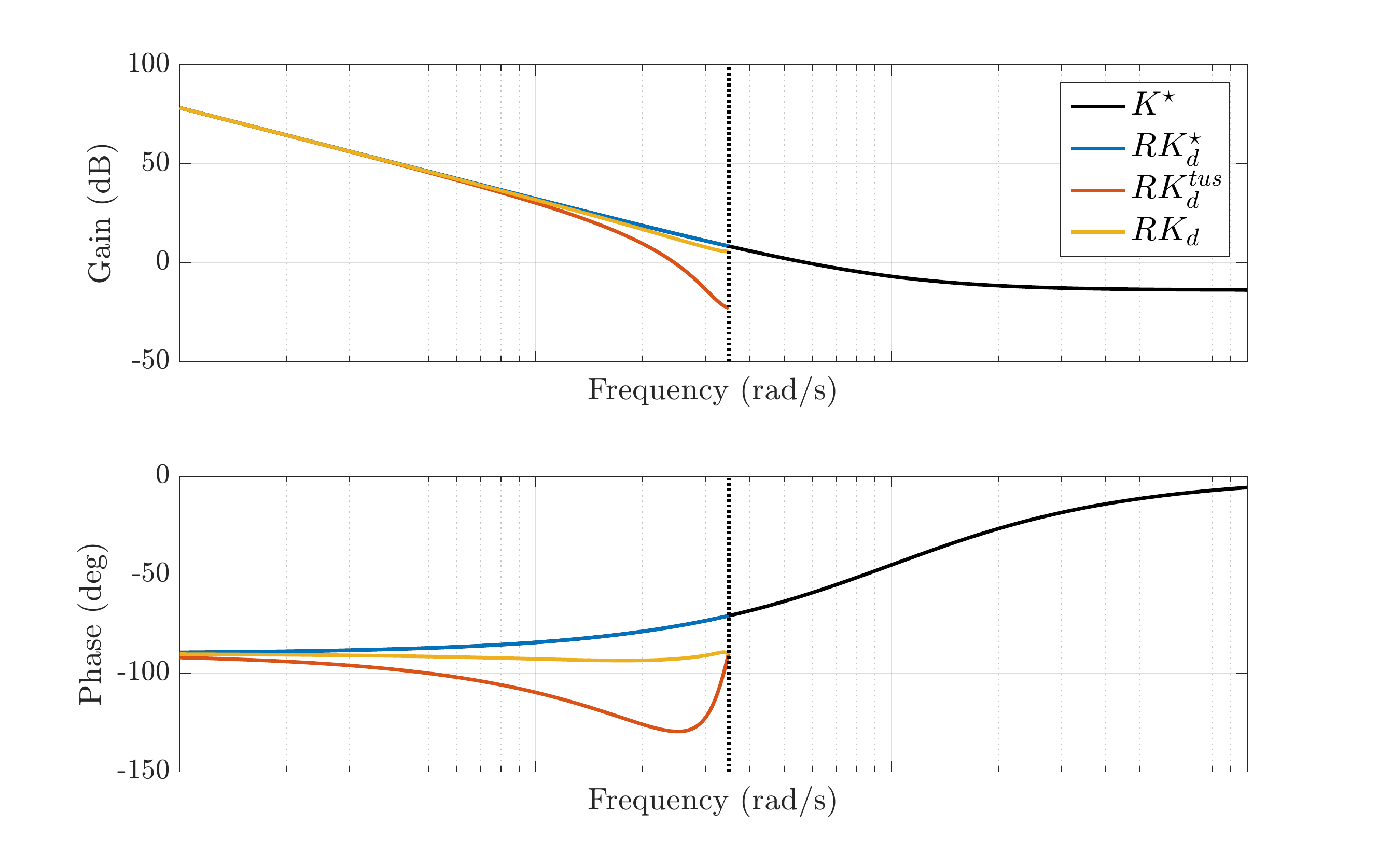}
    \caption{Frequency responses of the ideal continuous-time control-law $K^\star$ and its discrete-time counterparts completed with the response of the ideal holder for the DC-motor.}
    \label{fig:dcmotK}
\end{figure}

From a frequency-domain perspective, the frequency responses of the different control-law are reported in Figure \ref{fig:dcmotK}. First, one can notice that the ideal discrete-time control-law matches exactly the continuous-time one up to the Nyquist frequency (vertical dashed bar). The mismatch of the HLDDC controller $K_d$ with $K_d^\star$ comes from the stable projection but is still closer than the Tustin discretisation $K_d^{tus}$.

Note that for small enough sampling period $T$, both approaches leads to equivalent results. In that case, the advantage of the HLDDC mainly lies in the direct embedding of the discretisation step.

\subsection{Flexible transmission}
\begin{figure}
    \centering
    \includegraphics[width=\linewidth]{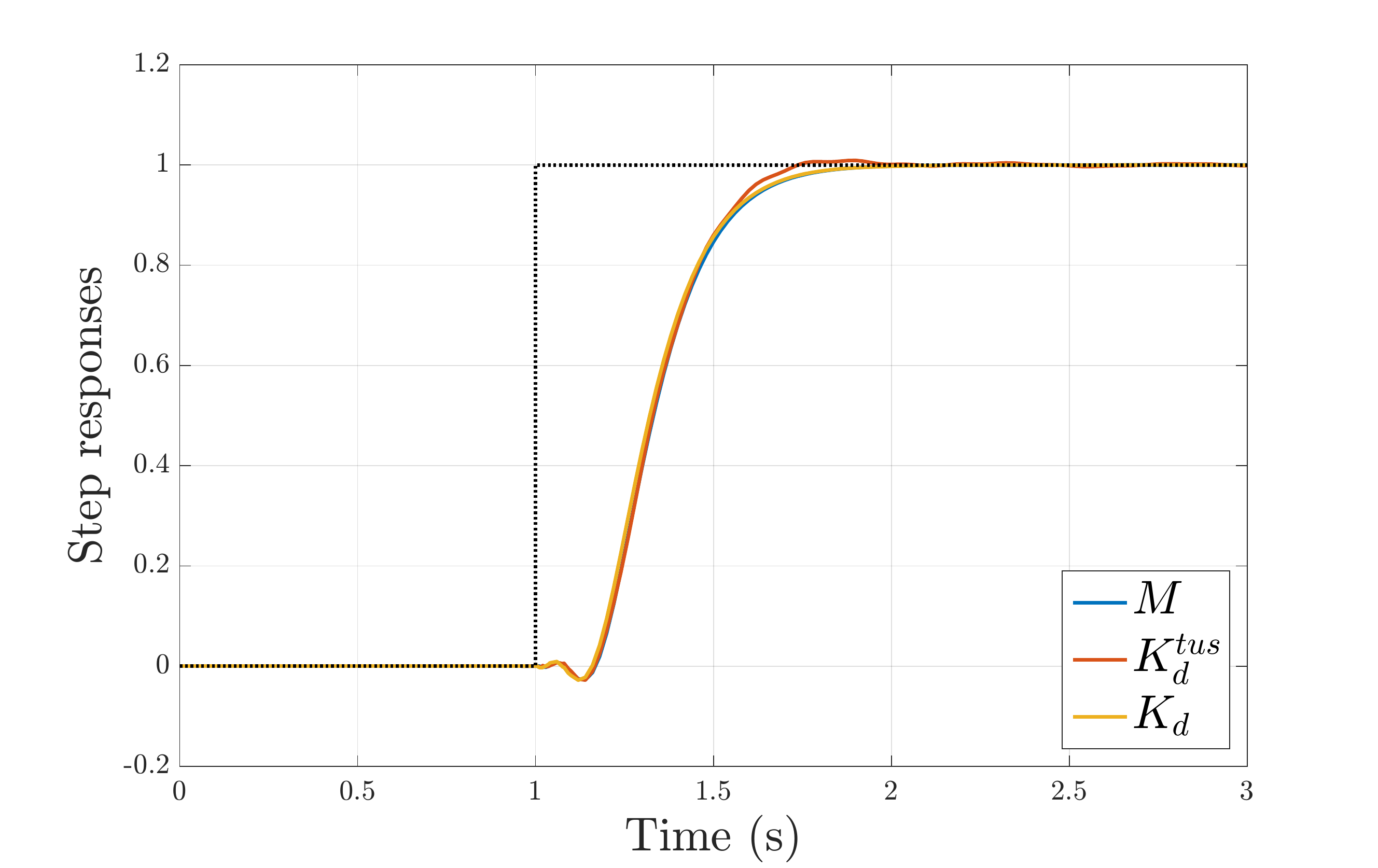}
    \caption{Step responses of the reference model and the closed-loops obtained with the direct approach and the separated discretisation of the control-law for the flexible transmission.}
    \label{fig:flextrans}
\end{figure}

\begin{figure}
    \centering
    \includegraphics[width=\linewidth]{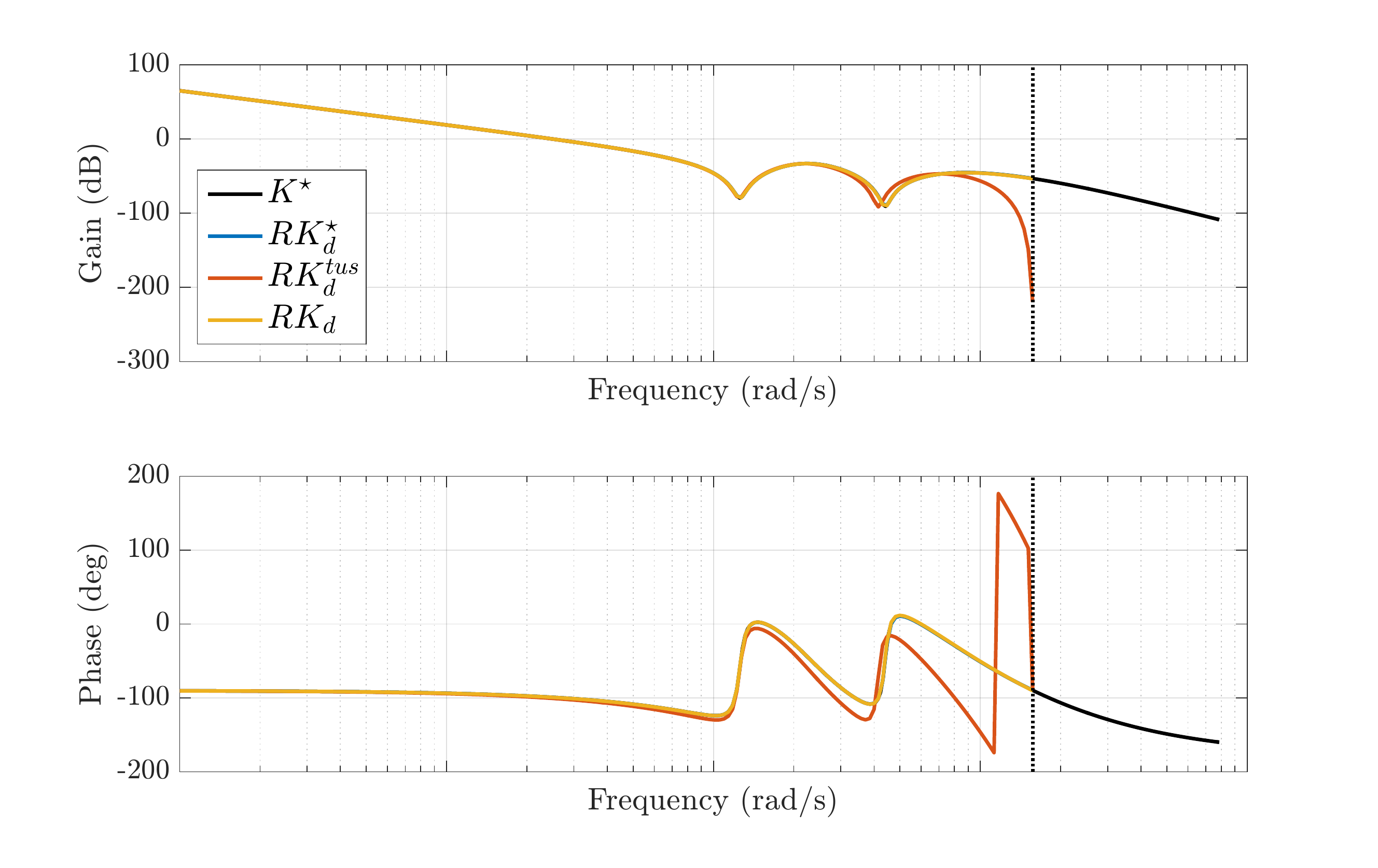}
    \caption{Frequency responses of the ideal continuous-time control-law $K^\star$ and its discrete-time counterparts completed with the response of the ideal holder for the flexible transmission.}
    \label{fig:flextransK}
\end{figure}

Here one considers the following plant,
\begin{equation}
P(s) = \frac{0.03616 (s-140.5) (s-40)^3}{ (s^2 + 1.071s + 157.9) (s^2 + 3.172s + 1936)}.
\end{equation}
As the plant is non-minimum phase, its closed-loop performances are limited and an arbitrary reference model may not be achievable while maintaining internal stability. In fact, the reference model must be selected with care so that the ideal controller $K^\star$ is stable and leads to internal stability of the closed-loop as detailed in \cite[chap. 4]{kergus:2019:phd}. The latter suggests a pre-treatment process to account for the performances limitations of the plant within the reference model, leading here to 
\begin{equation}
\scriptsize
M(s) = \frac{100 (s-140.6) (s-37.39) (s^2 - 82.6s + 1710)}{(s+10)^2 (s+37.39) (s+140.6) (s^2 + 82.6s + 1710)}.
\end{equation}
The same comparison as before is performed for the sampling period $T=0.02s$ leading to the following numerator/denominator coefficients for $K_d^{tus}$,
\begin{equation}
\small
    num(K_d^{tus}):\left [\begin{array}{r}
        0.0436 \\  -0.0949   \\ 0.0163  \\  0.1412  \\ -0.1515   \\ 0.0049   \\ 0.0837  \\ -0.0504  \\  0.0091
    \end{array}\right ],\,
    den(K_d^{tus}) : \left [\begin{array}{r}
        1.0000  \\  -3.6971 \\   5.6519 \\  -4.5649 \\   2.0349 \\  -0.4463  \\  0.0039  \\  0.0216\\   -0.0039

    \end{array} \right ]
\end{equation}
and for $K_d$,
\begin{equation}
\small
    num(K_d):\left [\begin{array}{r}0.1035\\   -0.0949 \\  -0.1132 \\   0.1837 \\  -0.0076 \\  -0.1134 \\   0.0292\\    0.0339    \\0.0033
    \end{array}\right ],\,
    den(K_d) : \left [\begin{array}{r} 1.0000  \\ -0.8853 \\  -1.4072  \\  1.4678 \\   0.4101 \\  -0.6735\\    0.0455 \\   0.0637 \\  -0.0212
    \end{array} \right ]
\end{equation}
The step responses are reported in Figure \ref{fig:flextrans}. In that case, the difference between the two approaches is barely noticeable. Larger differences can be observed in the frequency domain in Figure \ref{fig:flextransK} where the HLDDC approach matches exactly $K_d^\star$ while the Tustin controller leads to a larger mismatch.

However, in that case, the HLDDC approach is very sensitive to the sampling period and its behaviour drastically deteriorate for some value of $T$. This problem seems to be related to the determination of the minimal realisation order of the interpolant and its projection onto a stable subspace that induces, here, high errors w.r.t. the interpolation conditions \eqref{eq:kdjwinterp}.

\section{Conclusion}

This technical note describes a simple modification to the Loewer Data-Driven Control approach presented in \cite{kergus:2019:phd} so that it can directly create a discrete-time control-law instead of a continuous-time one thus embedding the usual a posteriori discretisation step.

This hybrid approach can be as (or even more) effective than the usual discretisation approach but it remains quite sensitive to some parameters like the sampling period. This point is currently under investigation.

\bibliographystyle{plain}
\bibliography{bibli}

\end{document}